\documentclass[runningheads]{llncs}
\usepackage{graphicx}
\usepackage{amsmath,amssymb} 
\usepackage{color}

\usepackage{booktabs}
\usepackage{nicefrac}
\usepackage[T1]{fontenc}
\usepackage[latin9]{inputenc}
\usepackage{array}
\usepackage{multirow}

\begin{document}
\pagestyle{headings}
\mainmatter

\def\ACCV20SubNumber{705}  

\title{Lossless Image Compression Using a Multi-Scale Progressive Statistical
Model} 
\titlerunning{Lossless Image Compr. Using a Multi-Scale Progressive Statistical Model}
%
\author{Honglei Zhang\inst{1}\orcidID{0000-0002-8229-852X} \and
Francesco Cricri\inst{1}\orcidID{0000-0002-1521-420X} \and
Hamed R. Tavakoli\inst{1}\orcidID{0000-0002-9466-9148} \and
Nannan Zou\inst{1,2}\orcidID{0000-0001-5553-5975} \and
Emre Aksu\inst{1}\orcidID{0000-0001-7363-2824} \and
Miska~M.~Hannuksela\inst{1}\orcidID{0000-0003-3405-0850}}

\authorrunning{H. Zhang et al.}
%
\institute{Nokia Technologies, Tampere, Finland \and
Tampere University, Finland \\
\email{\{honglei.1.zhang, francesco.cricri, hamed.rezazadegan\_tavakoli, \\
nannan.zou.ext, emre.aksu, miska.hannuksela\}@nokia.com}\\
\url{http://www.nokia.com}}

\maketitle

\begin{abstract}

Lossless image compression is an important technique for image storage
and transmission when information loss is not allowed. With the fast
development of deep learning techniques, deep neural networks have
been used in this field to achieve a higher compression rate. Methods
based on pixel-wise autoregressive statistical models have shown good
performance. However, the sequential processing way prevents these
methods to be used in practice. Recently, multi-scale autoregressive
models have been proposed to address this limitation. Multi-scale
approaches can use parallel computing systems efficiently and build
practical systems. Nevertheless, these approaches sacrifice compression
performance in exchange for speed. In this paper, we propose a multi-scale
progressive statistical model that takes advantage of the pixel-wise
approach and the multi-scale approach. We developed a flexible mechanism
where the processing order of the pixels can be adjusted easily. Our
proposed method outperforms the state-of-the-art lossless image compression
methods on two large benchmark datasets by a significant margin without
degrading the inference speed dramatically. 

\end{abstract}


\section{Introduction}

A lossless image compression system converts an input image into a
bitstream that is smaller in size and the original image can be fully
reconstructed. These systems bring several benefits in data storage,
manipulating, and transferring where information loss is not allowed.
On the other hand, lossy image compression aims at a much higher compression
rate by allowing the data to be reconstructed with a certain amount
of degradation. PNG \cite{libpnghome} and FLIF \cite{sneyers2016fliffree}
are image formats that support lossless image compression. Other image
compression techniques, such as JPEG 2000 \cite{jpeg}, WebP\cite{anew},
and HEVC \cite{sullivan2012overview} support both lossy and lossless
image compression.

A typical lossless image compression system contains a probability
model and an entropy codec. The probability model is built up from
a statistical model that gives an estimation of the statistical characteristics
of an input image. The entropy codec encodes/decodes symbols in the
input data into/from variable-length prefix-free codes according to
the statistics given by the probability model. According to Shannon$^{\prime}$s
source coding theorem \cite{shannon1948amathematical}, the optimal
code length of a symbol is $-\log_{2}P$, where $P$ is the probability
of the symbol to be encoded. Arithmetic coding is an efficient entropy
coding algorithm that has been adopted in many image/video coding
systems, such as HEVC \cite{sullivan2012overview}. 

Accurate probability estimation is the most critical aspect of a lossless
image compression system. Traditional lossless image compression system,
such as PNG \cite{libpnghome} uses a static statistical model with
Huffman coding to encode the difference of the pixels and the predicted
value based on its neighboring pixels. AVC \cite{marpe2006theh264mpeg4}
and HEVC \cite{sullivan2012overview} applies an adaptive model, known
as CABAC, to give a better estimation of the probabilities \cite{marpe2003contextbased}.
The CABAC model updates the model parameters during the process of
encoding/decoding according to the content of the image. With the
fast development of deep learning techniques, deep neural networks
have been used to achieve more accurate statistical estimations \cite{ho2020compression,townsend2019hilloclossless,johnston2017improved,hoogeboom2019integer,mentzer2020learning,cao2020lossless,mentzer2019practical}.
Unlike traditional image/video coding systems \cite{sullivan2012overview,marpe2006theh264mpeg4},
which encode residuals between the input symbols and the predicted
symbols, deep neural network-based image/video compression systems
normally use the estimated value distribution function directly, since
deep neural networks are more capable of modeling complicated distribution
functions. 

An autoregressive statistical model uses pixels that have already
been processed as a context to derive the value distribution function
of the pixels to be encoded/decoded \cite{oord2016conditional,oord2016pixelrecurrent,salimans2016pixelcnn}.
These models can achieve good compression rates. However, at the inference
stage, i.e. encode/decoding stage, the input data are processed sequentially
and an expensive deep neural network is involved to process every
pixel. Thus, the systems based on these models are only capable of
encoding/decoding small images \cite{mentzer2019practical}. To address
this problem, L3C \cite{mentzer2019practical} and SReC \cite{cao2020lossless}
adopt a multi-scale approach. Pixels at each scale are processed in
a parallel manner, which greatly speeds up the encoding/decoding procedure.
However, the compression rate is compromised significantly. In \cite{mentzer2020learning},
the authors proposed another type of context-based statistical model
where an input image is first compressed with a lossy compression
method and then used as the context to infer the value distribution
function of the pixels in an input image. 

Flow-based approach \cite{ho2020compression,hoogeboom2019integer,dinh2017density}
learns a invertible function that converts an input image into a latent
representation with a predefined distribution function. At the decoding
stage, the inverse function of the encoding function is applied to
reconstruct the input image. These algorithms also process pixels
in a parallel manner and have similar computation complexities for
encoding and decoding. However, the capacity of these methods is limited
by the constrain that the encoding and decoding functions must be
mutually invertible. Large neural networks have to be used to improve
the performance of the system. 

In this paper, we propose a lossless image compression system based
on a multi-scale progressive statistical model. The proposed method
significantly improves the compression rate without heavily degrading
the encoding/decoding speed. Our system incorporates a flexible framework
where the compression rate and encoding/decoding speed can be adjusted
easily. Using the proposed model, we have improved the state-of-the-art
compression rate on the ImageNet64 dataset and the OpenImage dataset
by a significant margin. 

\section{Related Work}

\subsection{Pixel-wise autoregressive statistical model \label{subsec:pixel-wise-autoregressive-statis}}

Pixel-wise autoregressive statistical models, such as PixelRNN \cite{oord2016pixelrecurrent},
PixelCNN \cite{oord2016conditional} and PixelCNN++ \cite{salimans2016pixelcnn},
model the joint distribution of the pixels in an image as the product
of the distribution of each pixel conditioned on the previous pixels.
Let $x$ be an image, $N$ be the number of pixels, and $x_{1},x_{2},\cdots,x_{N}$
be the pixels arranged in a certain order, for example, a raster scan
order. The joint distribution of $x$ is defined by 
\begin{equation}
p(x)=\prod_{i=1}^{N}p\left(x_{i}\vert x_{1},x_{2},\cdots,x_{i-1}\right),\label{eq:conditional_probability}
\end{equation}
where $p\left(x_{i}\vert x_{1},x_{2},\cdots,x_{i-1}\right)$ is modeled
by a deep neural network. During the training and encoding stage,
a masked convolutional neural network (CNN) is applied to ensure that
the probability estimation of the current pixel only depends on the
previous pixels. More importantly, with the masked CNN architecture
the system can be trained efficiently as a standard deep CNN. However,
at the decoding stage, pixels can only be processed sequentially.
This limitation prevents the pixel-wise autoregressive model from
being used in practice \cite{mentzer2019practical}. 

\subsection{Multi-scale autoregressive statistical model \label{subsec:Multi-scale-autoregressive-stati}}

Multi-scale autoregressive statistical models proposed in L3C \cite{mentzer2019practical}
and SReC \cite{cao2020lossless} model the joint distribution of $x$
as the product of the conditional distributions in multiple scales.
Let $M$ be the number of scales, and $x^{(i)}$ be the representation
at scale $i$. For simplicity, we use $x^{(0)}$ to denote the input
image. We have $x^{(i)}=E\left(x^{(i-1)}\right)$, where $E(\cdot)$
is a encoding function that performs a type of downsampling operation
to an input representation. Let $C^{(i)}=\left\{ x^{(i)},x^{(i+1)},\cdots,x^{(M)}\right\} $
be the set of representations from scale $i$ to scale $M$. The joint
distribution of $x^{(0)},x^{(1)},\cdots,x^{(M)}$ is defined by 
\begin{equation}
p\left(x^{(0)},x^{(1)},\cdots,x^{(M)}\right)=\left(\prod_{i=0}^{M-1}p\left(x^{(i)}\vert C^{(i+1)}\right)\right)p\left(x^{(M)}\right),\label{eq:multi-scale_distribution_model}
\end{equation}
where $p\left(x^{(M)}\right)$ is the distribution function of the
last scale and modeled by the assumption that the elements in $x^{(M)}$
are independent and uniform distributed. The conditional distribution
at scale $i$ is parameterized as

\begin{equation}
p\left(x^{(i)}\vert C^{(i+1)}\right)=p\left(x^{(i)}\vert y^{(i+1)}\right),\label{eq:multi_scale_decoder_function}
\end{equation}
where $y^{(i)}=D\left(y^{(i+1)},x^{(i)}\right)$, and $D(\cdot)$
is a decoder function implemented using a deep neural network. The
system is trained to optimize the cross entropy with respect to the
estimated probability distribution $p\left(x^{(0)}\vert y^{(1)}\right)$.

\subsection{Flow-based statistical model}

Flow-based generative method is another technique that has been recently
used for lossless image compression, for example, in LBB \cite{ho2020compression},
Integer Flows \cite{hoogeboom2019integer}, and Real-NVP \cite{dinh2017density}.
A flow-based method learns an invertible mapping function from the
image space to a latent space where the dependent variable follows
a predefined distribution function. Let the mapping function be $z=f(x)$
where $z$ is the dependent variable in the latent space with a predefined
distribution function $p_{Z}(z)$. The distribution function of input
$x$ is defined by 
\begin{equation}
p(x)=p_{Z}(f(x))\left|\det\left(\frac{\partial f(x)}{\partial x}\right)\right|,\label{eq:flow-based_distribution}
\end{equation}
where $\det(\cdot)$ is the determinant of a matrix. Because of the
determinant operation, the Jacobian of $f(x)$ at $x$ must have certain
forms, for example, be a diagonal or upper triangular matrix. In Real-NVP
\cite{dinh2017density}, Dinh et.al. proposed a invertible function
using coupling layers that are defined by 
\begin{align}
z_{1:d} & =x_{1:d}\\
z_{d+1:N} & =x_{d+1:N}\odot\exp\left(s\left(x_{1:d}\right)\right)+t\left(x_{1:d}\right),\label{eq:coupling_layers}
\end{align}
where $N$ is the dimension of the input variable, $s(\cdot)$ and
$t(\cdot)$ are scale and translation functions, and $\odot$ is element-wise
product. The coupling layer partition the variables in $x$ into two
sets $x_{1:d}$ and $x_{d+1:N}$. $x_{1:d}$ is directly mapped to
the corresponding output variables. $x_{d+1:N}$ is mapped to the
output variables using scale and translation factors that are derived
from $x_{1:d}$. 

Next, we will give a detailed description of our proposed statistical
model and our lossless image compression system.

\section{Multi-Scale Progressive Statistical Model}

\subsection{Statistical model\label{subsec:Statistical-model} }

Similar to multi-scale approaches \cite{cao2020lossless,mentzer2019practical},
we downsample an input image into a number of low-resolution representations.
Pixels in lower resolution representations are used as a context to
estimate the distribution function of the pixels in a higher resolution
representation. 

In L3C \cite{mentzer2019practical}, a representation $x^{(i+1)}$
is derived from representation $x^{(i)}$ using an encoder function
$E(\cdot)$. This design not only complicates the system but also
makes the training less efficient since the system becomes a very
deep neural network from the end-to-end point of view. Experiments
show that increasing the number of scales does not improve the performance
\cite{mentzer2019practical}. In SReC \cite{cao2020lossless}, the
encoder function is defined as an average pooling function. SReC can
be trained more efficiently than L3C. However, extra bits must be
used to encode round errors. Inspired by the coupling layer in Real-NVP
\cite{dinh2017density}, we simply take a subset of pixels in $x^{(i)}$
as $x^{(i+1)}$. With this simplification, the probability distribution
function $p\left(x^{(0)}\right)$ can be directly factorized as 
\begin{equation}
p\left(x^{(0)}\right)=\left(\prod_{i=0}^{M-1}p\left(x^{(i)}\vert C^{(i+1)}\right)\right)p\left(x^{(M)}\right)\label{eq:multis_scale_simplified}
\end{equation}

To further improve the compression rate without dramatically compromizing
the encoding/decoding speed, we partition the pixels at each scale
into groups, i.e. $x^{(i)}=g_{1}^{(i)}\cup g_{2}^{(i)}\cup\cdots\cup g_{B_{i}}^{(i)}$,
where $B_{i}$ is the number of groups at scale $i$. The details
of the grouping methods will be described in Section \ref{subsec:Pixel-grouping-methods}.
The groups are processed sequentially. Once all pixels in a group
are processed, they are added to the context to improve the estimation
accuracy of the pixels in the next group. Taking this grouping operation
into consideration, we can further factorize Eq. \ref{eq:multis_scale_simplified}
to 
\begin{equation}
p\left(x^{(i)}\vert C^{(i+1)}\right)=\prod_{j=1}^{B_{i}}p\left(g_{j}^{(i)}\vert g_{1}^{(i)},g_{2}^{(i)},\cdots,g_{j-1}^{(i)},C^{(i+1)}\right),\label{eq:multi_group_model}
\end{equation}
where $g_{j}^{(i)}$ is the pixel group $j$ at scale $i$, $B_{i}$
is the number of groups at scale $i$, and $i=0,1,\cdots,M-1$. Let
$G_{j}^{(i)}=\left\{ g_{1}^{(i)},g_{2}^{(i)},\cdots,g_{j-1}^{(i)},C^{(i+1)}\right\} $
be the context for group $g_{j}^{(i)}$. Assuming the pixels in a
group are conditional independent, we have the probability distribution
function 
\begin{equation}
p\left(g_{j}^{(i)}\vert G_{j}^{(i)}\right)=\prod_{k=1}^{N_{j}^{(i)}}p\left(x_{j,k}^{(i)}\vert G_{j}^{(i)}\right),\label{eq:prob_model_pixles}
\end{equation}
 where $x_{j,k}^{(i)}\in g_{j}^{(i)}$ is pixel $k$ in group $j$
at scale $i$, and $N_{j}^{(i)}$ is the number of pixels in group
$g_{j}^{(i)}$. 

Next, we use a mixture of logistic distributions to model $p\left(x_{j,k}^{(i)}\vert G_{j}^{(i)}\right)$
in a similar way as defined in PixelCNN++ \cite{salimans2016pixelcnn}.
The parameters of the mixture model $p\left(x_{j,k}^{(i)}\vert G_{j}^{(i)}\right)$
are derived from a function of context $G_{j}^{(i)}$ using a deep
neural network. Note that, in this model, the logistic mean of the
distribution function of a subpixel is modeled as a weighted sum of
the logistic means of previous subpixels. 

Given the statistical model defined by Eqs. \ref{eq:multis_scale_simplified},
\ref{eq:multi_group_model}, and \ref{eq:prob_model_pixles}, the
deep neural network is trained to minimize the cross entropy of the
input $x^{(0)}$, i.e. 
\begin{equation}
\mathbb{E}_{x^{(0)}\sim q\left(x\right)}\left[-\log p\left(x^{(0)}\right)\right],\label{eq:entropy_of_system}
\end{equation}
where $q(x)$ is the true distribution of the input image and $\mathbb{E}(\cdot)$
is the expectation of a random variable. 

\subsection{Lossless image compression system }

\begin{figure}
\begin{centering}
\begin{tabular}{c}
\includegraphics[width=12cm]{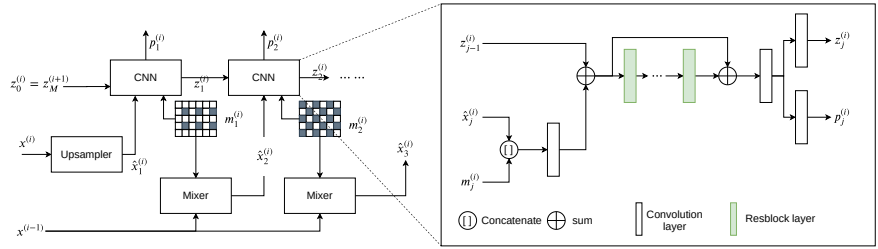}\tabularnewline
\end{tabular}
\par\end{centering}
\caption{The lossless image compression system using the proposed multi-scale
progressive statistical model. Yellow boxes indicate tensors. \label{fig:Progressive-autoregressive-stati} }
\end{figure}

Figure \ref{fig:Progressive-autoregressive-stati} illustrates our
lossless image compression system using the proposed multi-scale progressive
statistical model. For clarity, the figure only shows a system of
two scales. 

In this system, an input image $x^{(0)}$ is first downsampled to
low-resolution representations $x^{(1)}$ and $x^{(2)}$. As mentioned
in Section \ref{subsec:Statistical-model}, we take a subset of pixels
in $x^{(i)}$ as $x^{(i+1)}$. If the pixels are selected in a checkerboard
pattern, the system is equivalent to using the nearest neighbor downsampling
operation as the encoder function $E(\cdot)$ as used in \cite{cao2020lossless,mentzer2019practical}.
This simplification improves training efficiency and helps us to develop
a more flexible architecture to achieve better performance. With this
design, we can increase the number of scales without suffering from
the gradient vanishing/exploding problem that a very deep neural network
system may encounter. 

\begin{figure}
\begin{centering}
\begin{tabular}{c}
\includegraphics[width=12cm]{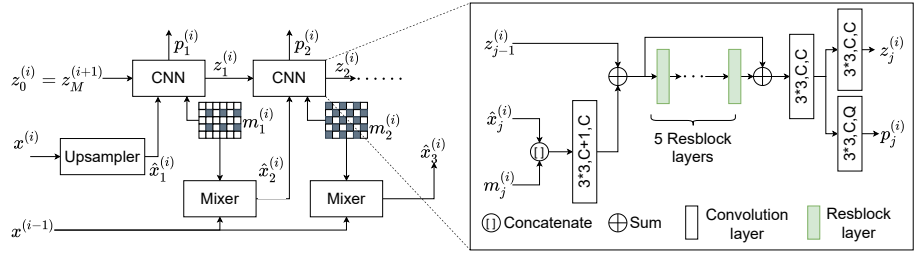}\tabularnewline
\end{tabular}
\par\end{centering}
\caption{The architecture of the progressive prediction model in Figure \ref{fig:Progressive-autoregressive-stati}.
The notations in convolution blocks are in the format of ``kernel\_size{*}kernel\_size,input\_channels,output\_channels''.
The Resblock layer is the basic Resnet block defined in \cite{he2015deepresidual}.
No batch normalization is used in the system. $C$ is the network
size and $Q$ is the number of parameters for the mixture of logistic
distributions model. \label{fig:pregressive_pred_model}}
\end{figure}

Figure \ref{fig:pregressive_pred_model} illustrates the progressive
prediction model at scale $i$, which is the core component of our
system. An input tensor $x^{(i)}$ is first upsampled to $\hat{x}_{1}^{(i)}$
to have the same size as tensor $x^{(i-1)}$. For simplicity, we choose
the nearest-neighbor upsampling method in our experiments. Next, $\hat{x}_{1}^{(i)}$
is partitioned into multiple groups as described in Section \ref{subsec:Statistical-model}.
For each group of pixels, a deep CNN is used to estimate the parameters
for the value distribution functions of the pixels in that group.
The detailed structure of this deep CNN is illustrated on the right
side of Figure \ref{fig:pregressive_pred_model}. The deep CNN takes
three inputs: $\hat{x}_{j}^{(i)}$ is the tensor that holds a mixture
of predicted and true values of pixels in $x^{(i-1)}$; $m_{j}^{(i)}$
is the binary mask that indicates the true values in $x_{j}^{(i)}$;
$z_{j-1}^{(i)}$ is the context information passed from the previous
stage. The initial context $z_{0}^{(i)}$ is set to be the output
context of scale $i+1$. For the last scale when there is no context
information available, we set the context to be zero, i.e. $z_{0}^{M}=0$.
The deep CNN outputs parameters $p_{j}^{(i)}$ for the estimated value
distribution functions of the pixels in group $j$, and the context
information $z_{j}^{(i)}$. 

At the training and encoding stage, where the ground truth $x^{(i-1)}$
is available, after the pixels in group $j$ have been processed,
the system sets the corresponding values in $\hat{x}_{j+1}^{(i)}$with
the true pixel values using the mixer component, and updates mask
$m_{j+1}^{(i)}$ accordingly. Then, $\hat{x}_{j+1}^{(i)}$, $m_{j+1}^{(i)}$
and $z_{j}^{(i)}$ are passed to the deep CNN to derive $p_{j+1}^{(i)}$
and $z_{j+1}^{(i)}$. The entropy of the pixels in group $j+1$ is
calculated using $p_{j+1}^{(i)}$ and $x^{(i)}$ at the training stage.
At the encoding stage, the pixels are encoded by the arithmetic encoder
according to the estimated value distribution function using parameters
$p_{j+1}^{(i)}$. At the decoding stage, the estimated value distribution
functions for pixels in group $j$ are used to decode the true values
from the bitstream using the arithmetic decoder. $\hat{x}_{j+1}^{(i)}$
and $m_{j+1}^{(i)}$are updated in the same way as the training and
encoding stage. This procedure continues until all $B^{(i)}$ groups
are processed.

\subsection{Pixel grouping methods\label{subsec:Pixel-grouping-methods}}

Using the binary mask $m_{j}^{(i)}$, one can easily set the number
of groups and the method to group the pixels. The grouping methods
determine the order of the pixels to be processed. In an extreme case
when $B^{(i)}$ is equal to the number of pixels to be processed at
scale $i$, the statistical model is equivalent to a pixel-wise autoregressive
model described in Section \ref{subsec:pixel-wise-autoregressive-statis}.
Next, we show some grouping methods that can be applied in the proposed
system. 

\subsubsection{Random grouping}

In this method, pixels to be processed at scale $i$ are randomly
assigned to $B^{(i)}$ groups. Note that the grouping is performed
in advance and agreed between the encoder and decoder so that the
pixels can be decoded correctly at the decoding stage. 

\subsubsection{Grouping with a fixed pattern}

With this method, the pixels are grouped by a predetermined fixed
pattern. Figure \ref{fig:grouping_fixed_pattern} shows two options
for the fixed pattern grouping. In (a), an input image is first partitioned
into 4x4 blocks and the pixels are grouped according to the index
number shown in the figure. Note that this grouping method is similar
to the architecture in \cite{cao2020lossless}. Since the value distribution
function are estimated using the available pixels in the neighboring
area as the context, it is intuitive to process the pixels that have
the most available pixels in its neighboring area. In (b), the pixels
are assigned to 6 groups in the order of the number of available neighboring
pixels. 

\begin{figure}
\begin{centering}
\begin{tabular}{ccc}
\begin{tabular}{|c|c|c|c|c|c|}
\hline 
1 & 2 & 1 & 2 & 1 & 2\tabularnewline
\hline 
3 & x & 3 & x & 3 & x\tabularnewline
\hline 
1 & 2 & 1 & 2 & 1 & 2\tabularnewline
\hline 
3 & x & 3 & x & 3 & x\tabularnewline
\hline 
\end{tabular} & \qquad{} & %
\begin{tabular}{|c|c|c|c|c|c|c|c|}
\hline 
5 & 3 & 2 & 4 & 5 & 3 & 2 & 4\tabularnewline
\hline 
6 & x & 1 & x & 6 & x & 1 & x\tabularnewline
\hline 
5 & 3 & 2 & 4 & 5 & 3 & 2 & 4\tabularnewline
\hline 
6 & x & 1 & x & 6 & x & 1 & x\tabularnewline
\hline 
\end{tabular}\tabularnewline
 &  & \tabularnewline
(a) &  & (b)\tabularnewline
\end{tabular}
\par\end{centering}
\caption{Two grouping methods with fixed patterns. (a) Grouping method (3 groups)
by a fixed pattern similar to \cite{cao2020lossless}; (b) Grouping
method (6 groups) according to the number of availability pixels in
the neighboring area. ``x'' indicates available pixels before grouping.
The number indicates the group to which the pixel belongs. Pixels
are processed group by group according to their group index number.
\label{fig:grouping_fixed_pattern}}
\end{figure}

\subsubsection{Dynamic grouping\label{subsec:Dynamic-grouping}}

Instead of using a predefined fixed pattern, one can also group the
pixels dynamically according to the content of the image. For example,
the grouping can be determined by the expected entropy values of the
pixels. As the progressive statistical model outputs the parameters
of the value distribution functions, the expected entropy values can
be calculated. However, this calculation is quite expensive. We chose
to use an upper bound to approximate the expected entropy values.
Let random variable $X$ follow a mixture of logistic distributions
determined by the number of mixtures $K$, mixture weights $\pi_{k}$,
means $\mu_{k}$ and scales $s_{k}$ where $k=1,2,\cdots,K$. An upper
bound of the expected entropy of $X$ can be defined by 
\begin{equation}
U(X)=\sum_{k=1}^{K}\pi_{k}\log\pi_{k}+\sum_{k=1}^{K}\pi_{k}\left(\ln s_{k}+2\right).\label{eq:entropy_upper_bound}
\end{equation}

The derivation of this upper bound can be found in the appendix. Given
the approximation of the expected entropy values defined by Eq. \ref{eq:entropy_upper_bound},
the pixels can be processed in an ascending order or descending order.
Our experiments have shown that the descending order yields a better
compression rate. In practice, this means that the system should process
the most difficult pixels first. This counter-intuitive behavior actually
helps to improve the overall compression rate since determining the
most difficult pixels can make other pixels more predictable. This
can be easily understood in the case of determining the borderline
of two homogeneous areas. 

\section{Experiments}

\subsection{Impact of grouping method}

We first evaluate the impact of the grouping methods described in
Section \ref{subsec:Pixel-grouping-methods}. In this experiment,
we trained the proposed probability model with different grouping
methods using the training split of the CIFAR-10 dataset \cite{krizhevsky2009learning}.
3-scale architecture is used. For the dynamic grouping method, we
partitioned the pixels into 3 groups at each scale according to the
estimated entropy upper bound calculated using Eq. \ref{eq:entropy_upper_bound}.
We used the Adam optimization method with a learning rate of 1E-4
and a batch size of 128. Since CIFAR-10 is a dataset of small images
(size 32x32), we trained the system until the training converged.
Table \ref{tab:Comparing-BPP-on} shows bits-per-pixel (BPP) results
on the validation split of the CIFAR-10 dataset. 

\begin{table}
\caption{Compression rate in bits-per-pixel on the validation split of the
CIFAR-10 dataset using different grouping methods described in Section
\ref{subsec:Pixel-grouping-methods}. \label{tab:Comparing-BPP-on}}

\centering{}%
\begin{tabular}{cccc}
\toprule 
Random Grouping & \quad{}Fixed Pattern (a) & \quad{}Fixed Pattern (b) & \quad{}Dynamic Grouping\tabularnewline
\midrule
10.83 & 10.66 & \textbf{10.45} & 10.60\tabularnewline
\bottomrule
\end{tabular}
\end{table}

Table \ref{tab:Comparing-BPP-on} shows that an ordered grouping method
performs better than the random grouping method. It also shows that
the dynamic grouping method described in Section \ref{subsec:Dynamic-grouping}
has a better compression rate comparing to the fixed pattern (a) in
Figure \ref{fig:grouping_fixed_pattern}. It should be noted that
the two methods have the same number of groups at each scale. According
to the results, increasing the number of groups can significantly
improve the compression rate as the fixed pattern (b), which contains
6 groups of pixels at each scale, outperforms all other grouping methods
(3 groups) with a significant margin. In our experiments, we also
noticed that the system with a fixed pattern grouping method converged
faster than the dynamic grouping method. 

\subsection{Impact of the number of scales}

In this experiment, we evaluated the impact of the number of scales
using the CIFAR-10 dataset. We use the fixed pattern (a) as the grouping
method. All other system parameters are the same except for the number
of scales. 

\begin{table}
\caption{Compression rate in bits-per-pixel on the validation split of the
CIFAR-10 dataset using different scales. \label{tab:impact_scales}}

\centering{}%
\begin{tabular}{cccc}
\toprule 
\quad{}2 scales\quad{} & \quad{}3 scales\quad{} & \quad{}4 scales\quad{} & \quad{}5 scales\quad{}\quad{}\tabularnewline
\midrule
10.95 & 10.66 & 10.614 & \textbf{10.612}\tabularnewline
\bottomrule
\end{tabular}
\end{table}

The results in Table \ref{tab:impact_scales} show that increasing
the number of scales can improve the compression rate. However, the
improvement from 4 scales to 5 scales is negligible. Note that the
tensor size at scale 4 accounts for less than 0.4\% of the input image.
Compressing this small amount of data does not bring much benefit
to the overall compression rate. 

\subsection{Compression performance on the ImageNet64 and the OpenImage dataset}

Next, we compare the proposed method against other state-of-the-art
methods on the two main datasets that have been used to benchmark
lossless image compression methods. The ImageNet64 dataset \cite{chrabaszcz2017adownsampled}
contains images from the ImageNet dataset \cite{deng2009imagenet}
that are downsampled to 64x64 pixels. The training split contains
1.28M images and the validation split contains 50K images. The OpenImage
dataset \cite{kuznetsova2020theopen} is a dataset with 9M images
with annotations for various computer vision tasks. For a fair comparison,
we prepared the training and the validation split of the OpenImage
dataset in the same way as \cite{mentzer2020learning,cao2020lossless,mentzer2019practical}.
PNG format of the OpenImage dataset was used to avoid JPEG compression
artifacts as suggested in \cite{cao2020lossless}. 

We define three profiles, as shown in Table \ref{tab:Three-profiles-used},
for our lossless image compression system and evaluated their performance
on the ImageNet64 and the OpenImage dataset. The difference between
the ``big'' and the ``normal'' profile is the grouping method.
Note that fixed pattern (b) has 6 groups at each scale while fixed
pattern (a) has 3 groups. The ``extra'' profile defines a bigger
system by increasing the number of scales from 3 to 4, the network
size (the number of channels used in convolution layers) from 64 to
128, and the number of mixtures of the mixture of logistic distributions
from 5 to 10. Note that the ``big'' profile does not increase the
network size compared to the ``normal'' profile. 

\begin{table}
\caption{The three profiles used to evaluate system performance.\label{tab:Three-profiles-used}}

\centering{}%
\begin{tabular}{c|ccc}
\hline 
 & normal  & big  & extra \tabularnewline
\hline 
scales & 3 & 3 & 4\tabularnewline
grouping method\enskip{} & \enskip{}fixed pattern (a)\enskip{} & \enskip{}fixed pattern (b)\enskip{} & \enskip{}fixed pattern (b)\tabularnewline
network size & 64 & 64 & 128\tabularnewline
number of mixtures & 5 & 5 & 10\tabularnewline
\hline 
\end{tabular}
\end{table}

The system is trained with Adam optimization and learning rate $2\times10^{-4}$
without weight decay. For the ImageNet64 dataset, the batch size is
set to 64. For the OpenImage dataset, we use randomly cropped 128x128
patches from input images as the input. The batch size for the ``normal''
profile is 64, the ``big'' profile is 32, and the ``extra'' profile
is 16. The learning rate is dropped by 10 times if a plateau is reached.
The ImageNet64 dataset is trained for 40 epochs and the OpenImage
dataset is trained for 20 epochs. All experiments had been executed
on an NVIDIA DGX-1 system with Tesla V100 GPUs. All training was performed
on a single GPU training mode. 

The validation split of the OpenImage dataset contains 500 images
randomly selected from the original validation set of the OpenImage
dataset. The selected images are the same as those used in \cite{mentzer2020learning,cao2020lossless,mentzer2019practical}.
If an input image is too large to be processed because of the limitation
of the GPU memory, we partition the image into patches and the bitstreams
of all patches are concatenated to generate the final output file.
The size of the patches is 496x496 for the ``normal'' and the ``big''
profile, and 256 for the ``extra'' profile. We note that this tiling
operation degrades the compression rate because of the border effects.
The reported compression rates of our methods are calculated from
the size of the final output file which includes also extra metadata
information such as image size and the number of scales. 

Table \ref{tab:Compression-results-on} shows the compression results
on the validation split of the ImageNet64 and the OpenImage dataset.
To be compared with other literature, both bits-per-pixel and bits-per-subpixel
(shown in parenthesis) are reported. The parameter column shows the
number of parameters for the neural network-based methods. 

\begin{table}
\caption{Compression results on the validation split of the ImageNet64 dataset
and the OpenImage dataset. The numbers under the datasets columns
are bits-per-pixel and bits-per-subpixel (in parenthesis). An empty
cell indicates that the field is not relevant and ``-'' symbol indicates
that the value is not reported by the authors. \label{tab:Compression-results-on}}

\centering{}%
\begin{tabular}{c|cccc}
\hline 
 & method & \enskip{}parameters\enskip{} & \enskip{}ImageNet64\enskip{} & \enskip{}OpenImage\enskip{}\enskip{}\tabularnewline
\hline 
\multirow{3}{*}{Traditional} & PNG \cite{cao2020lossless} &  & 17.22 (5.74) & 12.09 (4.03)\tabularnewline
 & WebP \cite{cao2020lossless} &  & 13.92 (4.64) & 9.09 (3.03)\tabularnewline
 & FLIF \cite{cao2020lossless} &  & 13.62 (4.54) & 8.61 (2.87)\tabularnewline
\hline 
\multirow{1}{*}{Flow-based} & IDF \cite{hoogeboom2019integer} & 84.33M & 11.70 (3.90) & 8.28 (2.76)\tabularnewline
\hline 
VAE-based & HiLLoC \cite{townsend2019hilloclossless} & - & 11.70 (3.90) & -\tabularnewline
\hline 
\multirow{6}{*}{Context-based\enskip{}} & RC \cite{mentzer2020learning} & - & - & 8.37 (2.79)\tabularnewline
 & L3C \cite{mentzer2019practical} & 5.01M & 13.26 (4.42) & 8.97 (2.99)\tabularnewline
 & SReC \cite{cao2020lossless} & 4.20M & 12.90 (4.29) & 8.10 (2.70)\tabularnewline
 & \enskip{}Ours (normal)\enskip{} & \textbf{1.87M} & 11.89 (3.96) & 8.14 (2.71)\tabularnewline
 & Ours (big) & \textbf{1.87M} & 11.78 (3.93) & 7.88 (2.63)\tabularnewline
 & Ours (extra) & 9.93M & \textbf{11.33 (3.78)} & \textbf{7.48 (2.49)}\tabularnewline
\hline 
\end{tabular}
\end{table}

The results in Table \ref{tab:Compression-results-on} show that our
methods significantly improve the compression rate on the ImageNet64
dataset with a much smaller model. The previous state-of-the-art method
achieved 11.70 BPP with a model of 84.33M parameters and the proposed
model achieved 11.33 BPP with a model of 9.93M parameters. On the
OpenImage validation dataset, the ``normal'' profile achieves the
same level of performance as the previous state-of-the-art methods
with a much smaller model. The ``big'' and ``extra'' profile of
the proposed method improves the state-of-the-art results by a big
margin. 

\subsection{Encoding/decoding time}

In this experiment, we evaluate the encoding and decoding time of
the proposed method versus other lossless image compression methods.
Methods that are based on pixel-wise autoregressive models are not
included in the investigation since it is well-known that these methods
are very slow and only capable of processing small images \cite{cao2020lossless,mentzer2019practical,ramachandran2017fastgeneration}.
For L3C, SReC, and our methods, we selected an input image of size
256x256 and compressed/decompressed the image using different methods
on the same system. Model loading time is not included in the measurement.
For the IDF method, we recorded the encoding time using a pretrained
CIFAR10 model on the CIFAR10 dataset. The LBB \cite{ho2020compression}
method is a flow-based method that requires auxiliary bits to be sent
together with the images to be encoded/decoded. It is included in
the comparison since it works similar to pixel-wise statistical models
that process pixels sequentially during the inference time. The values
of the HiLLoC and the LBB methods are calculated from the numbers
reported in \cite{townsend2019hilloclossless} and \cite{ho2020compression}
respectively. and We normalize the encoding/decoding time to be seconds
per 32x32 pixels since the reported numbers in literature are for
different sizes of images. Table \ref{tab:Compression/decompression-time-o}
shows the encoding and decoding time in seconds. Note that our implementation
is not optimized to achieve a fast encoding/decoding speed. 

\begin{table}
\caption{Encoding/decoding time in seconds per 32x32 pixels. \label{tab:Compression/decompression-time-o}}

\centering{}%
\begin{tabular}{>{\centering}p{1.3cm}>{\centering}p{1.2cm}>{\centering}p{1.2cm}>{\centering}p{1.3cm}>{\centering}p{1.2cm}>{\centering}p{1.2cm}>{\centering}p{1.3cm}>{\centering}p{1.3cm}>{\centering}p{1.3cm}}
\hline 
 & L3C \\
\cite{mentzer2019practical} & SReC \cite{cao2020lossless} & Ours (normal) & Ours (big) & Ours (extra) & HiLLoC \cite{townsend2019hilloclossless} & IDF\\
\cite{hoogeboom2019integer} & LBB \\
\cite{ho2020compression}\tabularnewline
\hline 
encoding\enskip{} & 0.0078 & 0.025 & 0.031 & 0.052 & 0.074 & 0.159 & 0.430 & 64.4\tabularnewline
decoding\enskip{} & 0.0070 & 0.025 & 0.029 & 0.049 & 0.070 & - & - & 65.9\tabularnewline
\hline 
\end{tabular}
\end{table}

As shown in Table \ref{tab:Compression/decompression-time-o}, the
proposed method has the same level of speed as SReC \cite{cao2020lossless}
with the ``normal'' profile. The ``big'' and the ``extra'' profile
show significant compression rate gains without heavily sacrificing
the inference speed. The results also show that the proposed methods
are also significantly faster than HiLLoC and IDF which also process
pixels in groups. Compared to pixel-wise methods like LBB, our proposed
methods have a tremendous advantage. 

\section{Discussions}

In this paper, we presented a multi-scale progressive statistical
model and a lossless image compression system based on it. The proposed
statistical model efficiently balances the accuracy that pixel-wise
models achieve and the speed that multi-scale models obtain. Experiments
show that the proposed system out-performs the state-of-the-art methods
by a significant margin on the two large benchmark datasets. The proposed
system achieves superior performance with smaller models compared
to other systems. The proposed system has a slightly higher computation
complexity compared to other ``fast'' methods with significant gains
in the compression rate. The proposed system incorporates a flexible
mechanism where the pixel grouping method can be specified easily.
We evaluated the static grouping methods using fixed patterns and
a dynamic grouping method based on the upper bound of the estimated
entropy. Experiments show that increasing the number of groups at
each scale significantly improves the compression rate. The dynamic
method shows certain gains compared to the static methods using fixed
patterns when the number of groups at each scale is the same. However,
the convergence speed at the training stage is much slower. This behavior
needs to be further studied and a mechanism that can speed up the
training with dynamic grouping methods shall be developed.



\bibliographystyle{splncs}
\bibliography{clean}

\end{document}